\documentclass[12pt,letterpaper]{article} 

\pdfoutput=1

\usepackage{amsmath}
\usepackage{booktabs}
\usepackage{multirow}
\usepackage{url}
\usepackage{cite}
\usepackage{caption}
\usepackage{graphicx}
\usepackage{authblk}
\newcommand{\tabref}[1]{Table~\ref{tab:#1}}
\newcommand{\figref}[1]{Figure~\ref{fig:#1}}

\title{A GPU Implementation for Two-Dimensional Shallow Water Modeling}

\author[1]{Kerry A. Seitz, Jr.\thanks{Corresponding author: kaseitz@ucdavis.edu}}
\author[1]{Alex Kennedy\thanks{Corresponding author: ajkennedy@ucdavis.edu}}
\author[2]{Owen Ransom}
\author[2]{Bassam A. Younis}
\author[3]{John D. Owens}
\affil[1]{Department of Computer Science}
\affil[2]{Department of Civil and Environmental Engineering}
\affil[3]{Department of Electrical and Computer Engineering\authorcr University of California, Davis}

\date{August 30, 2013}

\begin{document}
\maketitle

\section{Introduction}

Two dimensional water flow simulations are useful for engineering firms
to model floods, river/reservoir behavior, and dam break scenarios. These
models allow us to estimate the consequences of such events, which
can determine the allocation of funds towards high-risk areas. Much
of the work in flood modeling still uses one-dimensional solutions,
which model the river as a series of cross-sectional slices spanning
the entire breadth of the river. At each computational slice, they
simulate water elevation and velocity based on the nearby upstream and
downstream slices. The advantage of 1D solutions is that they vastly
simplify the computational difficulty of the problem. However, they
only model velocity in one direction per cross-sectional slice. This
method works well for large, steady rivers with low lateral velocity,
but these solutions are less apt at solving problems where water will
be flowing in multiple directions simultaneously.


To get more accurate results, we can instead use a two-dimensional
water model. These models require more computation because the domain
must be subdivided into cells in both the x- and y-direction to account
for flow velocities in multiple directions. Programs exist to solve
2D flow problems (such as SRH-2D), though most of them are solved on
the CPU~\cite{SRH2D}. These types of problems are massively parallel;
therefore, our goal is to develop a parallel implementation of a 2D
shallow water equation solver to improve runtime performance over current
serial implementations.

Our solution is taken directly from Owen Ransom's graduate thesis,
where he contrasts several numerical and discretization methods for
solving the shallow water equations~\cite{Ransom:Thesis}. Regarding his
solution, he says: ``the methods are based on approximate solutions of
partial differential equations of mass and momentum conservation for
unsteady free-surface flow of long waves." Owen provided us with his
serial Fortran code, which uses a 2D predictor-corrector MacCormack
method to solve for elevation, downstream velocity, and tangential
velocity for every time step.  In theory our solution is explicit,
so solving each time step is only dependent on a previous time step's
values. Practically, our 2D predictor/corrector adaptation requires us
to split one time step into 16 discrete sub-steps because each sub-step
requires neighboring cell information that was computed in the previous
sub-step (for the relevant mathematics of the model, refer to Ransom's
thesis~\cite{Ransom:Thesis}). However, within each of these 16 steps,
each cell's intermediate values can be computed independently of its
neighboring cells. Thus, our domain's cells can be solved in an arbitrary
order and our solution will remain deterministic. This cell-level
independence becomes important when we want to solve our problem in a
massively parallel manner.  With GPU parallelization schemes, there is no
guarantee that cell N will be solved before cell N+1; if cell N+1 relies
on cell N's computation, the solution becomes much less parallelizable.

We chose to parallelize our problem on the Graphics Processing Unit
(GPU) to take advantage of its massively parallel hardware.\footnote{We
will be using NVIDIA terminology to identify GPU hardware because they
are dominant in the scientific computations community. Other GPUs use
similar hardware but are named differently.} Modern consumer-grade GPUs
are designed with about 8-15 Streaming Multiprocessors (SMs), which act
as independent brains inside a GPU. Each SM can issue a vector operation
every clock-cycle, which executes the same math operation across
many pieces of data. These are known as Single Instruction Multiple
Data (SIMD) operations. Each piece of data is associated with a very
lightweight thread. Adjacent threads run on the same SM in groups called
``blocks." This mapping allows threads within a block to share a pool of
local, fast, ``shared" memory. GPUs do not guarantee the order in which
blocks and threads will be processed.  Fortunately, our water modeling
method is explicit, so it operates on each cell independently. Thus, the
order of execution is not important to the correctness of our solution.
Because our problem requires the same calculations to be performed
on many different pieces of data, this problem is well suited for
SIMD-style processors like the GPU. Therefore, we hope to achieve a
significant runtime speedup over CPU implementations using our GPU 2D
shallow water model.

\section{Related Works}
In recent years, several projects have developed 2D GPU-based solutions
for the shallow water equations. One such work by Brodtkorb et
al.\ uses a second-order Kurganov-Petrova scheme to calculate flux
across cell boundaries~\cite{Brodtkorb:2011}.  They suggest some
interesting improvements over the naive scheme, including an ``early
exit optimization", where they prematurely stop a block's computations
if it is completely dry. This optimization becomes particularly useful
for dam break datasets; many cells start dry, and early time steps need
to be short because of the rapidly moving water.

Ament et al.\ uses a version of the MacCormack method similar to ours to
solve for velocity and pressure~\cite{Ament:2011}. Their work focuses
largely on real-time visualization of fluid-flow. They use Lagrangian
Coherent Structures (LCS) with Finite-Time Lyapunov Exponents (FTLE) to
visualize a time-dependent vector field. The impressive part of their work
is that their simulation is fully interactive; they can dynamically change
the boundary conditions in their simulation and maintain a reasonable
15fps for computing the shallow water equations and visualizing them
over a 512x512 grid. They achieved this performance on three year old
hardware, so we are quickly reaching a point where modern hardware could
reproduce these results for a fully interactive screen-sized grid.

\section{Implementation}

We wrote our GPU implementation using NVIDIA's CUDA
platform~\cite{Nickolls:2008:CUDA}. This implementation consists of two
versions: a non-shared memory version and a shared memory version. Each
SM on a GPU contains shared memory with much faster access time than
the GPU's main/global memory. Therefore, a program that intelligently takes
advantage of shared memory provides some optimization opportunities.
We present an overview
of these two implementations in the sections below.

\subsection{Non-Shared Memory Version}

Our initial CUDA implementation consists of multiple kernels, each
of which must be executed for every time step. Kernel breaks provide
synchronization within a time step, which is necessary because each cell
requires intermediate data from neighboring cells (e.g. predictor values
when calculating the corrector value). We grouped threads into 1D thread
blocks, where each block operates on one row of data.

All data is stored in GPU global memory, so each thread must make multiple
global memory accesses to retrieve all of the data it needs. Because we
are operating on a 2D grid and retrieving adjacent cells simultaneously,
most memory accesses are coalesced.

Like many other explicit methods, Ransom's model uses a variable
time step. Calculating the next step size is an inherently serial
process, which is not well suited for GPU execution. Thus, we decided
to perform this computation on the CPU. This computation requires data
generated on the GPU, so we must copy this data to the CPU after each
time step. Memory copy overhead is significant, especially as the size
of the grid increases.\footnote{Discrete GPUs, such as those made by
NVIDIA and AMD, communicate with the CPU via the PCI Express bus. This
connection is much slower than the connection between the CPU and
the main Random Access Memory (RAM) on the computer.  Intel and AMD
are now shipping chips with a CPU and a GPU on a single die, so the
CPU and GPU can communicate efficiently using the main RAM.  However,
the GPUs on these chips are not as powerful as high-end discrete GPUs,
so there exists a trade-off between communication speed and processing
speed.} Running the step size calculation on the GPU resulted in a 10x
slowdown in overall runtime versus the version which uses the CPU. Thus,
we are willing to incur the memory transfer overhead to avoid performing
a serial calculation on the GPU.

\subsection{Shared Memory Version}

Within each thread block, data is reused multiple times -- a cell must
retrieve previously computed data not only from its own cell, but also
from its neighboring cells, most of which are also being operated
on within that block. Ideally, we only want to read each value from
global memory once.  Therefore, we can use shared memory to store local
copies of the values that will be accessed multiple times, which should
improve runtime performance. To further expose data reuse opportunities,
we grouped threads into two-dimensional blocks of size 16-by-16. Each
thread copies the data from the cell on which it is operating into
shared memory, and all other threads within the same block are free to
read that data directly from shared memory.

\begin{figure}[h]
\begin{centering}

\includegraphics[width=5in]{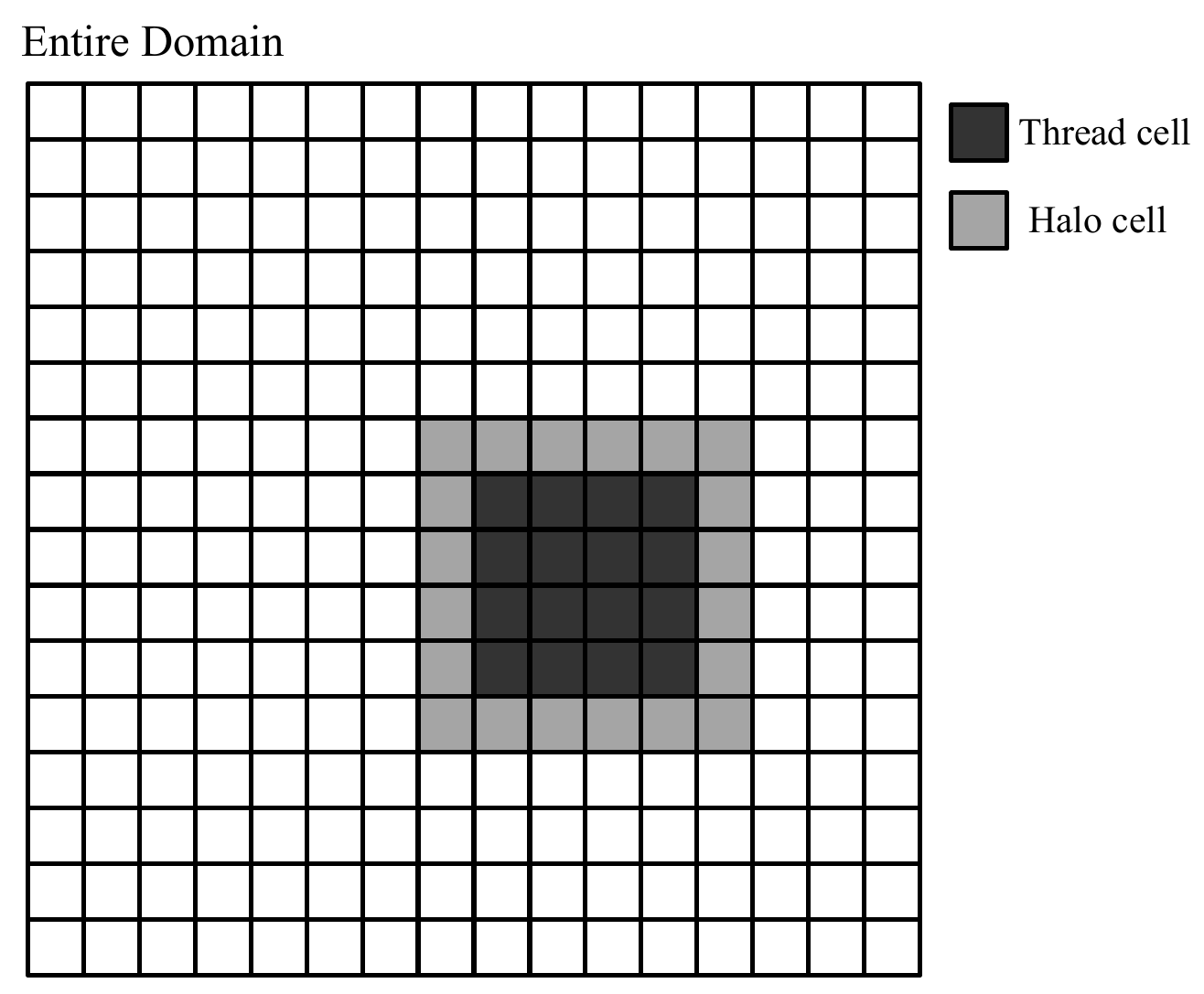}

\caption{A data halo representation. The ``thread cells'' are the cells
on which the current thread block are operating, and the ``halo cells''
are the extra data that the thread cells need to compute their values.}
\label{fig:halo}

\end{centering}
\end{figure}

Threads on the top, bottom, left, and right edges of a thread block
still require neighboring cell values. Because a thread block will not
be operating on, e.g., the north neighbors of the threads at the top
of the block, we must make special considerations to copy these outer
cells into shared memory. The extra data that must be copied into shared
memory is referred to as the ``halo'' (\figref{halo}). These halo values
will be copied into shared memory multiple times, likely on different
SMs. Even though this repeated copy is an inefficiency, the benefits of
using shared memory outweigh this downside.

\section{Results}

In order to evaluate the performance of our CUDA implementations, we
compared the runtime of the original Fortran code to the runtimes of our
shared and non-shared memory versions. We used five datasets to test our
code. These datasets varied in number of cells, length of simulation,
domain topology, and initial conditions. We ran these tests on a machine
running Ubuntu 12.04 with an Intel Core i7-3370K CPU and an NVIDIA GeForce
GTX 680 GPU. We used CUDA version 5.0 for our CUDA implementations. Our
performance results are presented in \tabref{results}.

\begin{table*}[h]
\begin{center}
\begin{tabular}{lccccc}
\toprule
							&										&											& \multicolumn{3}{c}{\textbf{Runtime}}\\\cmidrule{4-6}
							&										&											&	\multicolumn{1}{c}{}									&										& \textbf{CUDA}\\
							& 									& \textbf{Simulation}	& \multicolumn{1}{c}{}									& 		 							& \textbf{shared}\\
							& \textbf{Number}		& \textbf{End Time}		& \multicolumn{1}{c}{\textbf{Fortran}}	& \textbf{CUDA}			& \textbf{memory}\\
\textbf{Name}	& \textbf{of Cells}	& \textbf{(sec)}			& \multicolumn{1}{c}{\textbf{(sec)}}		& \textbf{(sec)}		& \textbf{(sec)}\\
\midrule
Five Drops 							& 40401			& 100		& 15.16					& 0.54		& 0.46\\
Inlet Flood							& 40401			& 1000	& 25.68$^\star$	& 6.21		& 5.69\\
Five Drops Big					& 1,048,576	& 100		& 1,115.67			& 10.33		& 9.13\\
Channel Flood$^\dagger$	& 1,048,576	& 1000	& N/A						& 57.02		& 48.26\\
Vortex$^\dagger$				& 1,048,576	& 1000	& N/A						& 352.07	& 295.35\\
\bottomrule
\end{tabular}

\caption{The performance results of the original Fortran code and our CUDA implementations.\\
\footnotesize
$^\star$The results from this test contained NaN values, indicating a
bug in the code.\\
$^\dagger$We created these two datasets in order to run more interesting tests using the CUDA code and to better observe the speedup of the shared-memory version over the non-shared memory version.  Thus, we did not use these datasets with the Fortran code.}

\label{tab:results}
\end{center}
\end{table*}

As shown, the CUDA code ran much faster than the Fortran code. The
degree to which the CUDA implementation outperformed the Fortran version
increases as the size of the domain increase, but not as the length of
the simulation increases. The reason for this behavior is that larger
grids require more parallelizable work whereas longer simulations require
more unparallelizable work (because the model requires synchronization
between time steps).

The shared memory version resulted in roughly a 10\% to 15\% speedup
over the non-shared memory version. We expect that this speedup would
be even greater with increased shared memory capacity because each SM
could execute more thread blocks in parallel. With more thread blocks,
an SM has more work it can execute in parallel and, thus, is better able
to hide latencies such as global memory accesses. However, each block
requires its own shared memory partition, so the shared memory capacity
limits the number of blocks an SM can run in parallel. We expect that
larger thread block sizes would also result in better performance because
we would get more data reuse within each block. However, current NVIDIA
GPUs are limited to 1024 threads per block, which limits our ability to
implement this improvement.

\section{Conclusion}

In this paper, we have presented a GPU implementation of Ransom's
two-dimensional shallow water model~\cite{Ransom:Thesis}. Our
implementation achieves a significant runtime speedup over our serial
test code.  This improvement allows researchers and engineers to study
water systems more efficiently and in greater detail. Our shared memory
optimization resulted in even better performance, and it has the potential
to scale linearly as the capacity of SM shared memory increases.

As demonstrated by our performance results, GPUs and other massively
parallel architectures have the capability to greatly impact scientific
research involving fluid modeling. By taking advantage of these devices,
engineers and scientists can improve their work by using more accurate,
larger scale models that would otherwise take prohibitively long to
simulate on the CPU.

\section{Future Work}
\subsection{Scaling the Solution}

Because each cell can be computed independently, our explicit method
demonstrates very good scalability. Our execution time has the desirable
property of scaling linearly with the size of our problem. Unfortunately,
this scalability is limited by a single GPU's computational and memory
resources.

One of our future goals is to distribute our domain's entire
work-load across multiple GPUs as efficiently as possible. Dividing
our computational domain requires careful and intelligent design to
maximize performance. Since the 16 kernel calls of each time step
must occur serially, distributing these kernels across multiple
GPUs is pointless. Dividing our problem so each of our N GPUs solves
$\frac{1}{N}$ of the time steps is equally unsuitable. However, we can
split our domain so each GPU solves one sub-grid per time step. We
expect a near-linear speedup as we increase the number of GPUs used
in our implementation. Redundant calculations, memory overhead, and
synchronization between devices will limit the actual speedup.

Each cell only relies on itself and a few neighboring cells to find
its solution from one time step to the next.  Because of this, each
GPU can work on one portion of our entire grid without interfering
with the others. This design introduces new problems and challenges;
cells on the edge of our domain require neighbors that belong to
a different GPU. Normally this would require device-to-device data
transfers and synchronizations to occur between each kernel.  However,
GPU programming guidelines suggest that this memory overhead is
expensive enough to justify additional computations instead.  To avoid
the costly synchronizations and memory transfers within each time step,
each GPU would do a minimal amount of duplicate work to calculate, e.g.,
their north and south neighbors' predictor/corrector values. With this
extra work, we can avoid the otherwise necessary synchronization points
within a single time step.  The best way to design our problem would
be to split our work into groups of horizontal scanlines. Each scanline
consists of multiple rows of our domain, and each of the N GPUs can work
on $\frac{1}{N}$ of the scanlines. This would allow our GPU-to-GPU memory
transfers to be coalesced, so we have minimal downtime between time steps.

\subsection{Cell Drying}

Our current solution is unable to handle dry cells. Because of this,
we had to carefully design our test-cases to make sure our computations
did not go unstable with negative water-elevations. This simplification
makes our code run faster than it otherwise would, but we suspect the
performance difference will be minimal. To tackle the cell-drying problem,
we will be working with Owen Ransom and other local civil engineers. This
is mandatory for our solution to become practical and useful, so it is
one of our top priorities.

\bibliography{finalPaper}{}
\bibliographystyle{plain}

\end{document}